\documentclass[12pt,preprint]{aastex}
\usepackage{times}
\usepackage{enumerate}
\usepackage{graphicx}
\usepackage{natbib}
\usepackage{epstopdf}
\usepackage{rotating}
\newcommand{\ltsima}{\stackrel{\textstyle <}{\sim}}
\newcommand{\simlt}{\scriptsize{\raisebox{-2pt}{$\ltsima$}}\normalsize}
\newcommand{\gtsima}{\stackrel{\textstyle >}{\sim}}
\newcommand{\simgt}{\scriptsize{\raisebox{-2pt}{$\gtsima$}}\normalsize}

\renewcommand{\thepage}{\rm\arabic{page}}

\newcommand{\water}{H$_2$O}

\newcommand{\waterice}{H$_2$O$_{\rm ice}$}
\newcommand{\ohice}{OH$_{\rm ice}$}
\newcommand{\oxy}{O$_2$}

\newcommand{\ico}{$^{13}$CO}

\newcommand{\asec}{$^{\prime\prime}$}
\newcommand{\amin}{$^{\prime}$}
\newcommand{\ddeg}{$^{\rm o}$}
\newcommand{\mh}{H$_2$}
\newcommand{\nh}{$n${(H$_2$)}}

\newcommand{\av}{{\em A}$_{\rm V}$}
\newcommand{\um}{$\:\mu$m}
\newcommand{\kms}{~km~s$^{{-1}}$}
\newcommand{\cmc}{cm$^{-3}$}
\newcommand{\cms}{cm$^{-2}$}

\newcommand{\go}{$G_{\rm o}$}
\newcommand{\dash}{$\,$--$\,$}

\newcommand{\tro}{3$_3\,$--$\,$1$_2$}
\newcommand{\trt}{5$_4\,$--$\,$3$_4$}
\newcommand{\kks}{K$\:$km s$^{-1}$}
\newcommand{\tgas}{$T_{\rm gas}$}
\newcommand{\ywat}{$Y_{H_2O}$}

\newcommand{\sigh}{$\sigma_H$}
\newcommand{\rmin}{$a_{\rm min}$}
\newcommand{\rmax}{$a_{\rm max}$}

\newcommand{\etal}{et$\,$al.}
\newcommand{\ti}{$\,\times\,$}

\renewcommand{\apj}{{\em Ap.~J.}}

\renewcommand{\apjs}{{\em Ap.~J.~Suppl.}}

\renewcommand{\mnras}{{\em M.N.R.A.S.}}

\renewcommand{\aap}{{\em A\&A}}
\renewcommand{\aj}{{\em A.~J.}}

\makeatletter
\def\fnum@figure{{Fig.~\thefigure}}
\long\def\@makecaption#1#2{
 \vskip 10pt 
 \setbox\@tempboxa\hbox{#1. #2}
 \ifdim \wd\@tempboxa >\hsize \unhbox\@tempboxa\par \else \hbox
to\hsize{\hfil\box\@tempboxa\hfil} 
 \fi}

\makeatother

\makeatletter
\def\ps@myheadings{\let\@mkboth\@gobbletwo
\def\@oddhead{\hbox{}\sl\rightmark \hfil \thepage}%
\def\@oddfoot{}\def\@evenhead{\thepage\hfil\sl\leftmark\hbox {}}%
\def\@evenfoot{}\def\sectionmark##1{}\def\subsectionmark##1{}}
\makeatother

\makeatletter
\def\subsection{\@startsection{subsection}{2}{\z@}{-3.25ex plus -1ex minus 
   -.2ex}{1.5ex plus .2ex}{ \em \rm}}
\makeatother

\newcounter{ncount}

\shorttitle{Search for O$_2$ Toward the Orion Bar}
\shortauthors{Melnick et al.}

\begin{document}

\begin{center}
{\large\bf {Herschel}$^*$ Search for O$_2$ Toward the Orion Bar} \\*[9mm]
Gary J. Melnick$^1$, Volker Tolls$^1$, Paul F. Goldsmith$^2$, Michael J. Kaufman$^3$, \\
David J. Hollenbach$^4$, John H. Black$^5$, Pierre Encrenaz$^6$,
Edith Falgarone$^7$, Maryvonne Gerin$^7$, \AA ke Hjalmarson$^5$,
Di Li$^8$, Dariusz C. Lis$^9$, Ren\'{e} Liseau$^5$,
David A. Neufeld$^{10}$, Laurent Pagani$^{6}$, \\
Ronald L. Snell$^{11}$, Floris van der Tak$^{12}$, and Ewine F. van Dishoeck$^{13, 14}$

\vspace{27mm}

%Rev.~3\\*[2mm]
%{\em \today} \\*[11mm]

\vspace{2.0in}

Received$\;$\rule{1.65in}{0.25mm}$\,$;~~~~~Accepted$\;$\rule{1.65in}{0.25mm}
\end{center}

\vspace{2.6in}

\noindent $^*$~Herschel is an ESA space observatory with science instruments provided by European-led
\phantom{0.}Principal Investigator consortia and with important participation from NASA

\clearpage

\setcounter{page}{2}

\pagestyle{myheadings}
\markright{{}\hfill{\hbox{\hss\rm Page }}}

\begin{enumerate}
\item Harvard-Smithsonian Center for Astrophysics, 60 Garden Street, MS 66, Cambridge, MA 02138, USA \\*[-6mm]

\item Jet Propulsion Laboratory, California Institute of Technology, 4800 Oak Grove Drive, Pasadena, 
CA 91109, USA \\*[-6mm]

\item Department of Physics and Astronomy, San Jo\'{s}e State University, San Jose, CA 95192, USA \\*[-6mm]

\item SETI Institute, Mountain View, CA 94043, USA \\*[-6mm]

\item Department of Earth \& Space Sciences, Chalmers University of Technology, Onsala Space Observatory, SE-439 92 Onsala, Sweden \\*[-6mm]

\item LERMA \& UMR8112 du CNRS, Observatoire de Paris, 61 Av. de l'Observatoire, 75014 Paris,
France \\*[-6mm]

\item LRA/LERMA, CNRS, UMR8112, Observatoire de Paris \& \'{E}cole Normale Sup\'{e}rieure, 24 rue Lhomond, 75231 Paris Cedex 05, France \\*[-6mm]

\item National Astronomical Observatories, Chinese Academy of Sciences, A20 Datun Road, Chaoyang District, 
Beijing 100012, China \\*[-6mm]

\item California Institute of Technology, Cahill Center for Astronomy and Astrophysics 301-17, 
Pasadena, CA 91125, USA \\*[-6mm]

\item Department of Physics and Astronomy, Johns Hopkins University, 3400 North Charles Street, 
Baltimore, MD 21218, USA \\*[-6mm]

\item Department of Astronomy, University of Massachusetts, Amherst, MA 01003, USA \\*[-6mm]

\item SRON Netherlands Institute for Space Research, P.O. Box 800, 9700 AV, and Kapteyn Astronomical Institute,
University of Groningen, Groningen, The Netherlands \\*[-6mm]

\item Leiden Observatory, Leiden University, P.O. Box 9513, 2300 RA, Leiden, The Netherlands \\*[-6mm]

\item Max-Planck-Institut f¬ur Extraterrestrische Physik, Giessenbachstrasse 1, 85748, Garching, Germany  \\*[-6mm]
\end{enumerate}

%\begin{center}
%ABSTRACT
%\end{center}

\renewcommand{\baselinestretch}{1.5}

\begin{abstract}
We report the results of a search for molecular oxygen (\oxy) toward the Orion Bar, a prominent
photodissociation region at the southern edge of the H$\,$II region created by the luminous Trapezium stars.
We observed the spectral region around the frequency of the \oxy\ $\rm N_J =\;$\tro\ transition at 487~GHz and 
the \trt\ transition at 774~GHz
using the Heterodyne Instrument for the Far Infrared on the {\em Herschel Space Observatory}.
Neither line was detected, but the 3$\sigma$ upper limits established here translate to a total 
line-of-sight \oxy\ column density
%less than 
$<\:$1.5$\,\times\,$10$^{16}$~\cms\ for an emitting region whose temperature is between 30$\:$K
and 250$\:$K, or
%or less than 
$\,<\;$1$\,\times\,$10$^{16}$~\cms\ if the \oxy\ emitting region is primarily at a 
temperature of $\simlt\,$100$\:$K.
%if the \oxy\ emitting region is primarily at a temperature of $\simlt\,$100$\:$K,
%the 3$\sigma$ upper limit to the total \oxy\ column density is 1$\,\times\,$10$^{16}$~\cms.  
Because the Orion Bar is oriented nearly edge-on relative to our line of sight, the observed column
density is enhanced by a factor estimated to be between 4 and 20 relative to the face-on value.  
Our upper limits imply that the face-on \oxy\ column density is less than 
4$\,\times\,$10$^{15}$~\cms, a value that is below, and possibly well below, model predictions
for gas with a density of 10$^4$\dash 10$^5$~\cmc\ exposed to
a far ultraviolet flux 10$^4$ times the local value, conditions inferred from previous observations of the Orion Bar.
The discrepancy might be resolved if: (1) the adsorption energy of O
atoms to ice is greater than 800$\:$K; 
%(2) \oxy\ emission fills less than 14\asec\ of the 28.2\asec\ HIFI beam at 774~GHz; 
(2) the total face-on \av\ of the Bar is less
than required for \oxy\ to reach peak abundance; (3) the \oxy\ emission arises within dense clumps
with a small beam filling factor;
or, (4) the face-on
depth into the Bar where \oxy\ reaches its peak abundance, which is density dependent, corresponds to
a sky position different from that sampled by our {\em Herschel} beams.

\end{abstract}


\renewcommand{\baselinestretch}{1.0}

\vspace{4.5in}

\baselineskip=24pt

\clearpage

\section{INTRODUCTION}

Searches for interstellar \oxy\ have a long history, but their motivation has evolved with time. 
Prior to the late-1990's, 
efforts to detect \oxy\ were driven largely by a desire to confirm its predicted role as a major reservoir
of elemental oxygen within dense molecular clouds and as the most important gas coolant -- after CO -- of cold
($T\,\simlt\,$30$\:$K), modestly dense (\nh$\,\simeq\,$10$^3$\dash 10$^4$~\cmc) gas
\citep*[cf.][]{Goldsmith78,Neufeld95}.  The launch of the {\em Submillimeter Wave Astronomy
Satellite (SWAS)} in 1998 and {\em Odin} in 2001, and the subsequent failure of these observatories
to detect \oxy\ toward a large number of sources at levels of a few percent of the abundances predicted
by equilibrium gas-phase chemical models, have forced a shift in
emphasis to a re-examination of the 
oxygen chemistry in dense molecular gas.  Today, interest in \oxy\ no longer lies in its being a significant
reservoir of elemental oxygen or in its cooling power.  Instead, because the abundance of gas-phase \oxy\
is set by a balance of various formation, destruction, and depletion processes thought to affect the
broader chemistry in dense gas -- such as gas-phase reactions, grain-surface reactions,
thermal sublimation, far-ultraviolet (FUV) photodesorption, cosmic-ray desorption, photodissociation,
and freeze out -- measures of
\oxy\ have become an important test of our current understanding of the relative effectiveness of these
processes.

The capabilities of the {\em Herschel Space Observatory's} Heterodyne Instrument for the Far-Infrared 
\citep[HIFI;][]{deGraauw10} have enabled improved searches for \oxy\ through: (1) its high sensitivity, including
at 487~GHz -- the frequency of the $\rm N_J =\;$3$_3$\dash 1$_2$ transition observed previously by {\em SWAS}
and {\em Odin}; and, (2) its broad frequency coverage that permits observations of additional \oxy\ 
submillimeter transitions, some of which are expected to exhibit stronger emission than the \tro\ line
under certain physical conditions.  The Open Time Key Program ``Herschel Oxygen Project" 
(HOP; Co-PI's P. Goldsmith and R. Liseau) is designed to survey Galactic sources with the goal to
detect \oxy\ or set meaningful limits on its abundance within these regions.
Because the effectiveness of the processes that determine the
\oxy\ column density depends upon the gas density, temperature, and incident FUV
flux \go\ (scaling factor in multiples of the average Habing local interstellar radiation field; \citealt{Habing68}) 
among other parameters, testing these 
models requires that the HOP observations include a
range of source types, such as dense quiescent clouds, outflows and shocked gas regions, and FUV-illuminated
cloud surfaces \citep[see, for example,][]{Goldsmith11, Liseau12}.

In this paper, we report the results of a deep search for \oxy\ emission toward the Orion Bar, a
well known ionization front located approximately 2\amin\ southeast of the Trapezium stars in Orion at the
interface of the H$\,$II region created by these stars and the dense gas associated with the surrounding Orion
molecular cloud.  The Orion Bar lends itself well to the study of FUV-illuminated molecular gas
for several reasons, including its nearly edge-on geometry, its proximity
\citep[$\sim\,$420 pc;][]{Menten07, Hirota07, Kim08}, its relatively high density
(\nh$\,\simgt\,$3$\times$10$^4$~\cmc), and the strong (\go$\,\simeq\,$10$^4$\dash 10$^5$)
external FUV field irradiating this gas.   The Orion Bar, and sources like it, are of
particular interest since the dust grains within these regions are predicted to be sufficiently warm that
the thermal evaporation of O atoms from the grain surfaces is enhanced, resulting in a higher fraction
of O in the gas phase and the
increased production of \oxy\ via gas-phase chemical reactions (O$\,+\,$OH$\,\rightarrow\,$O$_2\,+\,$H). 
Under such circumstances,
the \oxy\ column density can be more than a factor of 10 greater than within gas exposed to
lower (i.e., \go$\,<\,$500) external FUV fields \citep[cf.][]{Hollenbach09}.  The inclusion of the Orion Bar
within the HOP program was intended to test this prediction.

The observations and data reduction methods are described in \S2 below. In \S3, we
present the resultant spectra and the upper limits to the \oxy\ integrated intensity.  In \S4, we 
review the excitation conditions within the Orion Bar and the
derived limits on the line-of-sight \oxy\ column density.  In \S5, we discuss these limits in the context of 
recent chemical models that trace the \oxy\ abundance from the FUV-illuminated cloud surface to the
deep interior.

\vspace{2mm}

\section{OBSERVATIONS AND DATA REDUCTION}

The {\em Herschel} HIFI observations presented here were carried out using the HIFI Band 1a receiver for
the \tro\ 487 GHz observations and the HIFI Band 2b receiver for the \trt\ 774 GHz observations.  
The 487 GHz observations
were conducted on operational day (OD) 291 in spectral scan dual beam switch (DBS) mode, while the
774~GHz observations were conducted on OD 297 in spectral scan DBS mode and on OD~509
in HIFI single point DBS mode.
Eight LO-settings were used for both the 487 GHz and the 774 GHz spectral scans to enable 
the spectral deconvolution, and the additional eight single point 774 GHz observations were 
observed also using eight different LO settings. The total integration time (on-source$\,+\,$off-source)
{\em for each polarization} was 0.93 hours for the 487 GHz spectral scan, 0.86 hours for the 774 GHz spectral
scan, and a total of 4.6 hours for the eight single point 774 GHz observations.  The 
full-width-at-half-maximum (FWHM) beam sizes were 44.7\asec\
at 487 GHz and 28.2\asec\ at 774 GHz.

\begin{figure}[t]
\centering
\vspace{-0.70in}
$\!$\includegraphics[scale=1.14]{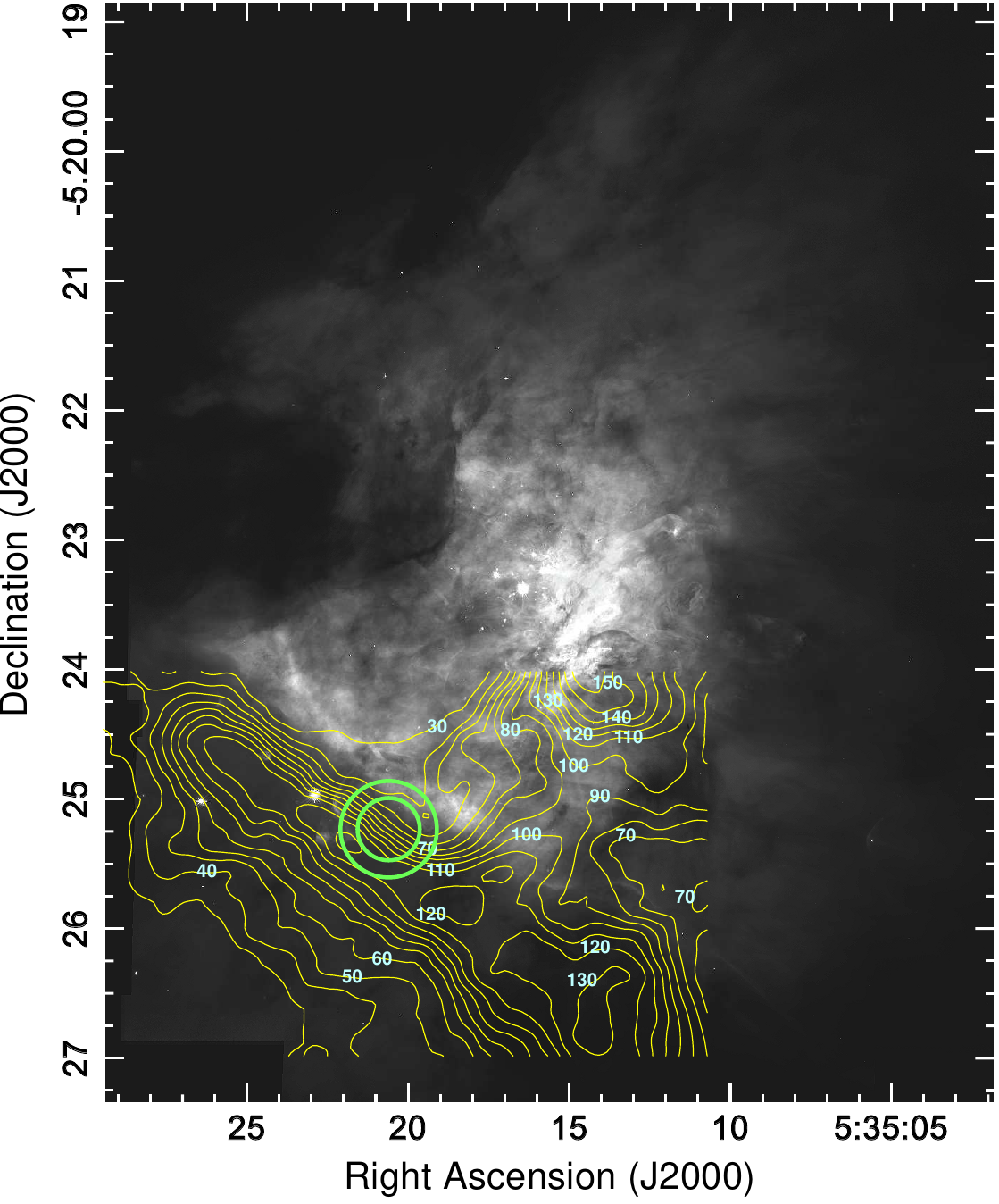}
%$\!$\includegraphics[scale=1.14]{fig1.eps}
\renewcommand{\baselinestretch}{0.99}
\vspace{3mm}
\caption{Position of the HIFI 44.7\asec\ and 28.2\asec\ beams at 487 GHz and 774 GHz, respectively,
superposed on an HST image of the Orion nebula \citep{ODell96}. 
Also shown are contours of \ico\ $J =\,$3\dash 2 integrated intensity for a portion of a larger map obtained
by \cite{Lis03}, with intensities in \kks\ noted.
The HIFI beams are centered at $\alpha\,=\,$05$^{\rm h}$35$^{\rm m}$20$^{\rm s}\!\!.$6,
$\delta\:=\,-$05\ddeg 25\amin 14\asec\ (J2000), toward the surface layers of the FUV-illuminated
Orion Bar where the \oxy\ emission is predicted to peak.}
\renewcommand{\baselinestretch}{1.0}
\label{finderchart}
\end{figure}

The observed position, $\alpha =\,$5h 35m 20.6s, 
$\delta =\,-$5\ddeg$\:$25\amin$\:$14.0Õ\asec\ (J2000), is shown in Fig.~\ref{finderchart}.
We applied the total observing time allotted to HOP observations of the Orion Bar to
a single spatial position -- versus multiple positions -- in order to achieve the lowest radiometric noise and, thus, 
the greatest sensitivity to weak \oxy\ emission.  In the absence of prior information about the possible \oxy\
spatial distribution, our choice of sky position was guided by the desire to place the 487~GHz and
774~GHz beam centers a distance corresponding to approximately 8 visual magnitudes into
the molecular gas measured from the ionization front, in accord with model predictions (see \S5 for a full discussion).
For an \mh\ density between 5\ti 10$^4$~\cmc\ and 5\ti 10$^5$~\cmc, applicable to
the interclump medium in the Bar, and \go$\,\simeq\,$10$^4$,
this corresponds to a projected angular distance of between 2.4\asec\ and 24\asec\ from the ionization
front.  As shown in Fig.~\ref{finderchart}, the selected position places the beams in the center of this
range, while the beam sizes encompass the full range.  The sky position parallel to the Orion Bar was 
selected to coincide with the molecular gas, as delineated by the \ico\ $J\,=\:$3\dash 2 emission
(see Fig.~\ref{finderchart}), and, for future analysis, one of the positions under present study by 
another {\em Herschel} Key Program.

The data were processed using the standard HIFI pipeline software HIPE version 7.3 \citep{Ott10},
spurious signals (``spurs") removed, spectra defringed, spectral scans deconvolved, and all data finally exported to
GILDAS-CLASS format.  Further processing was performed only on the Wide Band Spectrometer
(WBS) spectra (0.5 MHz channel spacing and 1.1 MHz effective spectral resolution) using the
IRAM GILDAS software package (http://iram.fr/IRAMFR/GILDAS/), including first-order baseline
removal, averaging of the 774 GHz spectral scans and frequency-aligned single point observations,
averaging of the H- and V-polarization spectra, and production of separate averages for both frequencies
and both sidebands.  The frequencies 
for the line identification were extracted from the JPL and CDMS databases \citep{Pickett98, Muller05}
as well as \cite{Drouin10} in the case of \oxy.

\vspace{2mm}

\section{RESULTS}

\begin{figure}[t]
\centering
\vspace{-0.30in}
%$\!$\includegraphics[scale=0.605]{fig1.eps}
$\!$\includegraphics[scale=0.58]{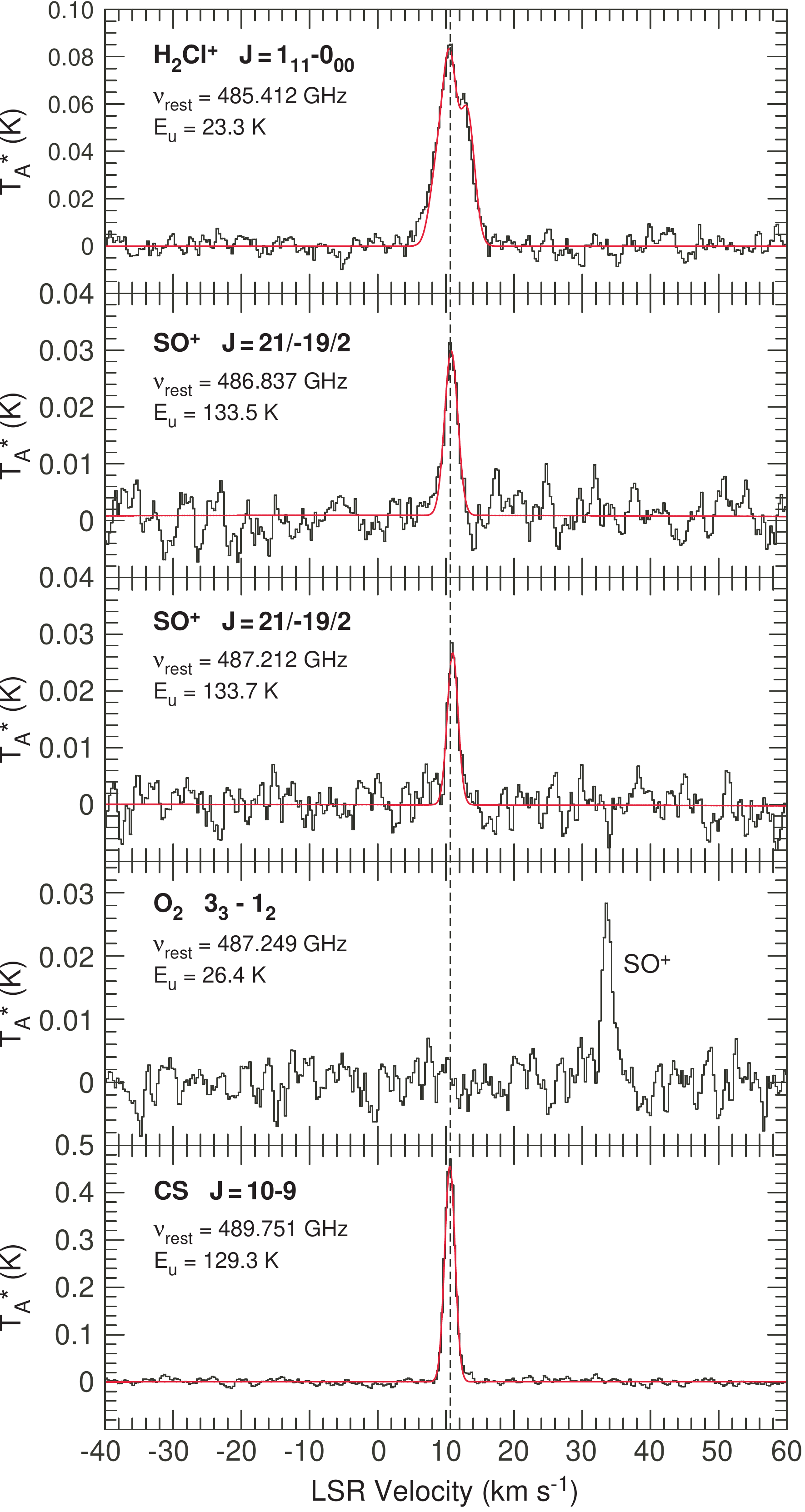}
\renewcommand{\baselinestretch}{0.97}
\vspace{0.1mm}
\caption{Averaged H and V polarization spectra obtained in HIFI Band 1a toward the Orion Bar, 
ordered by rest frequency, 
with the Gaussian fits given in Table 1 superposed in red.
Also indicated is the energy of the upper level for each transition, in Kelvins.  The H$_2$Cl$^+$
line is a blend of three, partially resolved, hyperfine components (see Table~1). An LSR
velocity of 10.7\kms\ is denoted with a vertical dashed line.}
\renewcommand{\baselinestretch}{1.0}
\label{Band1aspectra}
\end{figure}

\begin{figure}[t]
\centering
\vspace{-0.30in}
%$\!$\includegraphics[scale=0.605]{fig1.eps}
$\!$\includegraphics[scale=0.58]{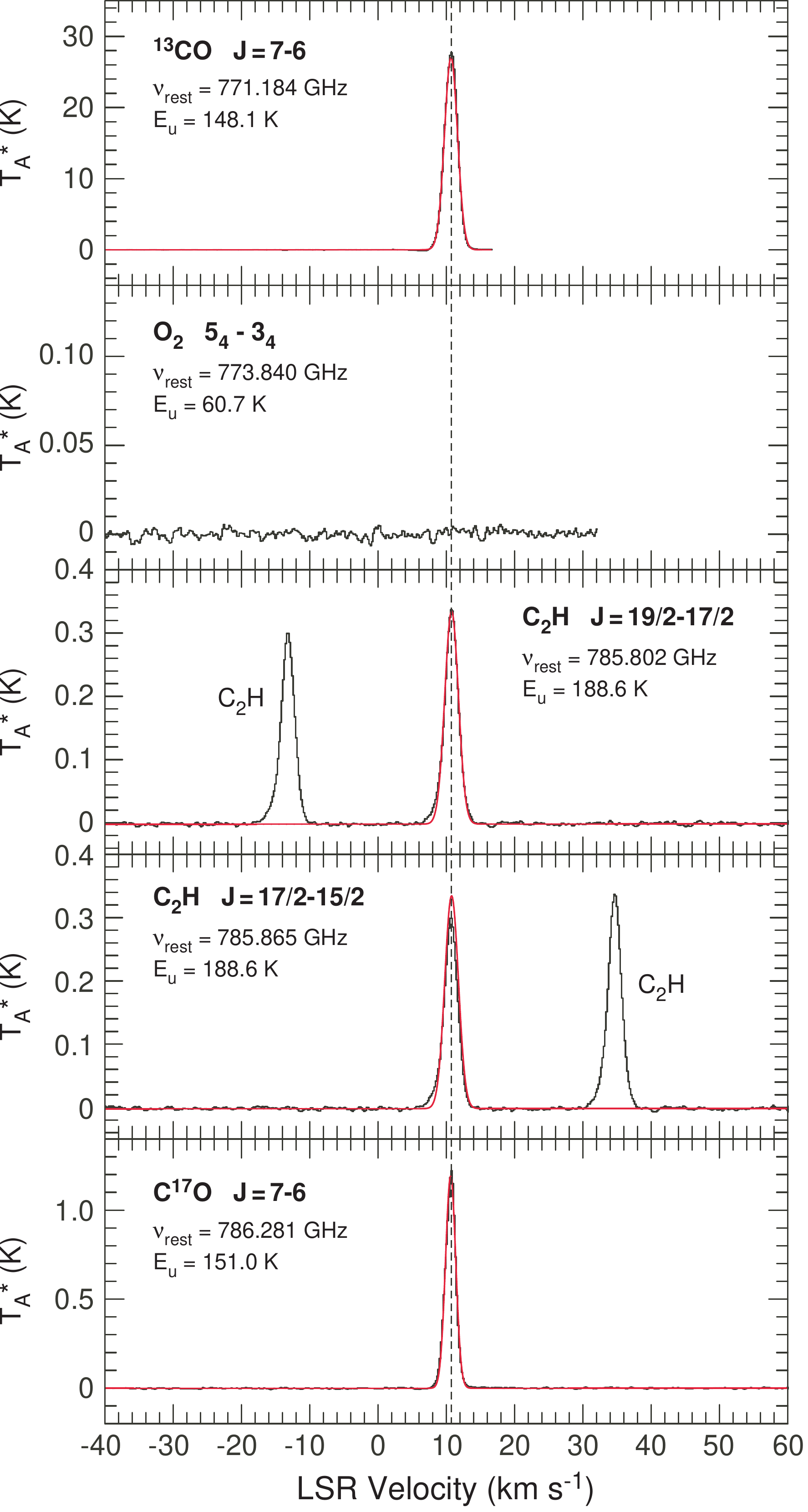}
\renewcommand{\baselinestretch}{0.97}
\vspace{-0.8mm}
\caption{Averaged H and V polarization spectra obtained in HIFI Band 2b toward the Orion Bar, 
ordered by rest frequency, 
with the Gaussian fits given in Table 1 superposed in red.
Also indicated is the energy of the upper level for each transition, in Kelvins.  An LSR
velocity of 10.7\kms\ is denoted with a vertical dashed line.  The frequency of the \ico\ $J=\,$7\dash 6
transition was near the edge of the band and, thus, the spectrum does not cover the full LSR velocity 
range of the other lines.  The \oxy\ spectrum has been truncated to remove features in the other sideband.}
\renewcommand{\baselinestretch}{1.0}
\label{Band2bspectra}
\end{figure}

A summary of the identified lines in the HIFI Band 1a and Band 2b spectra along with the
observing modes, integration times, and Gaussian fit parameters is provided in Table~1.
The summed H+V polarization spectra observed in Band 1a are shown in Fig.~\ref{Band1aspectra},
while those observed in Band 2b are shown in Fig.~\ref{Band2bspectra}.  With the exception
of the H$_2$Cl$^+$ chloronium 485~GHz spectrum, which is a blend of three hyperfine
components \citep[cf.][]{Lis10, Neufeld11}, all of the detected lines appear well fit by single Gaussian
profiles with a common LSR line center of 10.68$\,\pm\,$0.14\kms\ (1$\sigma$) and individual best-fit
FWHM line widths ranging from about 1.8\kms\ to 2.5\kms.

The upper limit to the integrated intensity of the \oxy\ \tro\ and \trt\ transitions is derived
assuming each line is described by a single Gaussian profile, as is the case for the other unblended lines we
detect toward this position.  The rms noise in the
\oxy\ \tro\ 487~GHz spectrum between LSR velocities of $-$110\kms\ and $+$25\kms\ -- a velocity range
within which there is no evidence for any spectral features -- is
2.62$\,$mK per 0.5 MHz channel.  Similarly, the rms noise in the \oxy\ \trt\ 774~GHz spectrum
between LSR velocities of $-$70\kms\ and $+$30\kms\ is 2.19$\,$mK per 0.5 MHz channel.
The intrinsic \oxy\ line widths along this line
of sight are unknown; however, we assume they
lie between the extremes of 1.8\kms\ and 2.5\kms\ (FWHM) measured for
the other unblended lines we detect along this line of sight (see Table~1).  This leads to 3$\sigma$ upper limits
of between 0.0150 and 0.0209 \kks\ for the \tro\ 487 GHz line and between 0.0126
and 0.0175 \kks\ for the \trt\ 774 GHz line.

\vspace{2mm}

\section{EXCITATION AND LIMITS ON THE O$_2$ COLUMN DENSITY}

The Orion Bar, like many other photodissociation regions (PDRs), displays emission from a variety of
ionic, atomic, and molecular
species best fit by a mix of gas densities and temperatures.  The broad
picture to emerge is that of a layer consisting of at least two components: interclump gas with \nh\
$\sim\,$3\dash 20$\,\times\,$10$^{4}\;$\cmc\ \citep{Hogerheijde95, Wyrowski97, Simon97, Marconi98}
surrounding clumps with \nh\ $\sim\,$10$^6$\dash 10$^{7}\;$\cmc\
\citep{Lis03, Owl00}, which comprise about 10\% of the mass \citep{Jansen95}.  
Gas temperature estimates similarly vary, depending on the species observed and
the component giving rise to most of the emission.  Within the denser well-shielded gas, the gas temperature
is thought to range between $\sim\,$50 and 85$\,$K \citep{Hogerheijde95, Gorti02}.
The gas temperature associated with the interclump medium
is estimated to be 85$\,\pm\,$30$\:$K \citep{Hogerheijde95}, with some gas temperatures associated with
the surfaces (\av$\,\simlt\,$1) of the denser clumps ranging as high as 220$\,$K
\citep{Jansen95, Batrla03, Goicoechea11}.  There is evidence for an even warmer component (300\dash 700$\,$K)
based on emission from pure rotational lines of \mh\ and far-infrared fine-structure lines of 
[O$\,$I] at 63 and 145\um\ and [C$\,$II] at 158\um\ \citep{Herrmann97, Allers05}.  This warmer component
is believed to arise in the gas between the ionization front and the molecular region traced by
\ico\ emission \citep{Walmsley00}.
The strength of the FUV field incident on the Orion Bar has been estimated to be 
\go$\,\simeq\,$1\dash 4$\times$10$^4$ based upon the total radiation from the Trapezium stars -- and
the O star $\theta^1$ Ori C in particular --
the intensity of the far-infrared [C$\,$II] and [O$\,$I] fine-structure lines mapped toward the 
Orion molecular ridge, the strength of several near-infrared lines whose intensities have been 
ascribed to recombinations to highly excited states of CI, and the strength of near-infrared NI lines 
excited by the fluorescence of UV lines \citep{Herrmann97, Marconi98, Walmsley00}.  
Given a density of $\sim\,$10$^5$~\cmc\ for the bulk of the material
and a \go\ of $\sim\,$10$^4$, models predict that the \oxy\ abundance peaks at \av$\:\simgt\:$8 mag.
\citep[cf.][]{Sternberg95, Hollenbach09}.   At these depths into the cloud, the gas temperature is predicted to be
30\dash 40$\:$K \citep{Hollenbach09}.  Thus, in our analysis, we consider the
possibility that the \oxy\ emission could arise in gas with temperatures anywhere between 30$\:$K and 250$\:$K.

The weak line flux of the \oxy\ magnetic dipole transitions makes them highly likely
to be optically thin.  Under the assumption that the \oxy\ emission uniformly fills the HIFI beam, 
the observed integrated intensity in a given transition is:

\vspace{-4.5mm}

\begin{equation}
\int T_{\rm mb}\,dv~=~\frac{h c^3}{8 \pi \nu^2 k}\;\,
A_{\rm u\ell}\:N({\rm O_2})\:f_u~=~5.15 \times 10^{-4}\;\frac{A_{\rm u\ell}\:N({\rm O_2})\:f_u}
{\nu_{\rm GHz}^2}~~~{\rm (K\;km\:s^{-1})}~,~
\end{equation}

\vspace{4mm}

\noindent where $T_{\rm mb}$ is the main beam temperature, $\nu$ is the
line frequency (and $\nu_{\rm GHz}$ is the line frequency in GHz), 
$A_{\rm u\ell}$ is the spontaneous decay rate between the transition upper level, $u$,
and lower level, $\ell$, $N$(\oxy) is the total \oxy\ column density in \cms, and $f_u$ is the fractional
population in the transition upper level.  The conversion between main beam and antenna temperature
makes use of the efficiencies reported in \cite{Roelfsema12}.

\begin{figure}[t]
\centering
\vspace{-0.20in}
%$\!$\includegraphics[scale=0.605]{fig1.eps}
$\!$\includegraphics[scale=0.77]{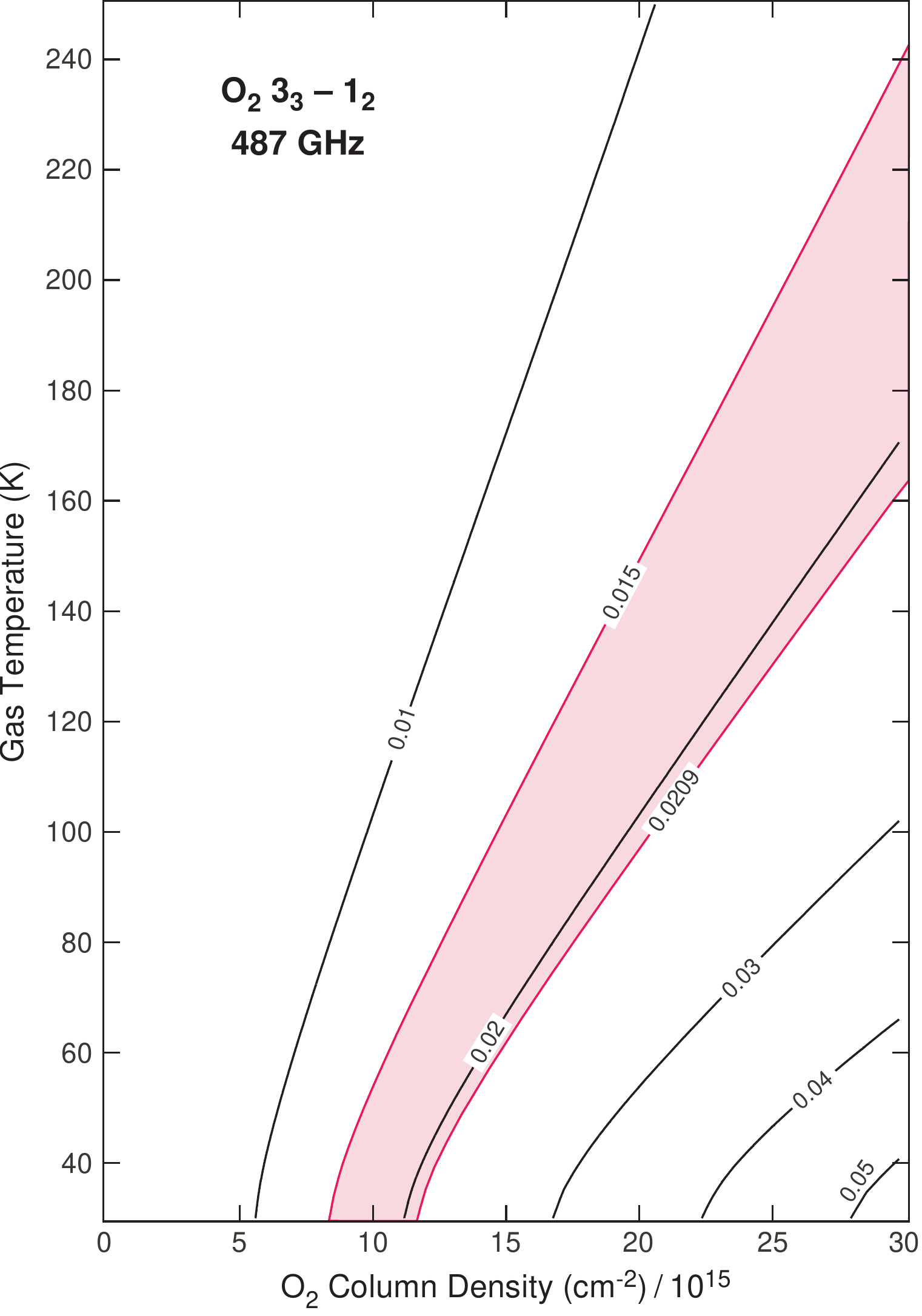}
\renewcommand{\baselinestretch}{0.99}
\vspace{3mm}
\caption{Contours of integrated antenna temperature, in K km s$^{-1}$, under a Gaussian line profile
versus line-of-sight \oxy\ column density and gas temperature assuming an aperture efficiency of 0.7.  
The shaded area bounds the range of upper limits to the \oxy\ 487 GHz integrated intensity assuming the
intrinsic \oxy\ line FWHM is 1.8\kms\ (0.0150 \kks) or 2.5\kms\ (0.0209 \kks).}
\renewcommand{\baselinestretch}{1.0}
\label{contours487}
\end{figure}

\begin{figure}[t]
\centering
\vspace{-0.20in}
%$\!$\includegraphics[scale=0.605]{fig1.eps}
$\!$\includegraphics[scale=0.77]{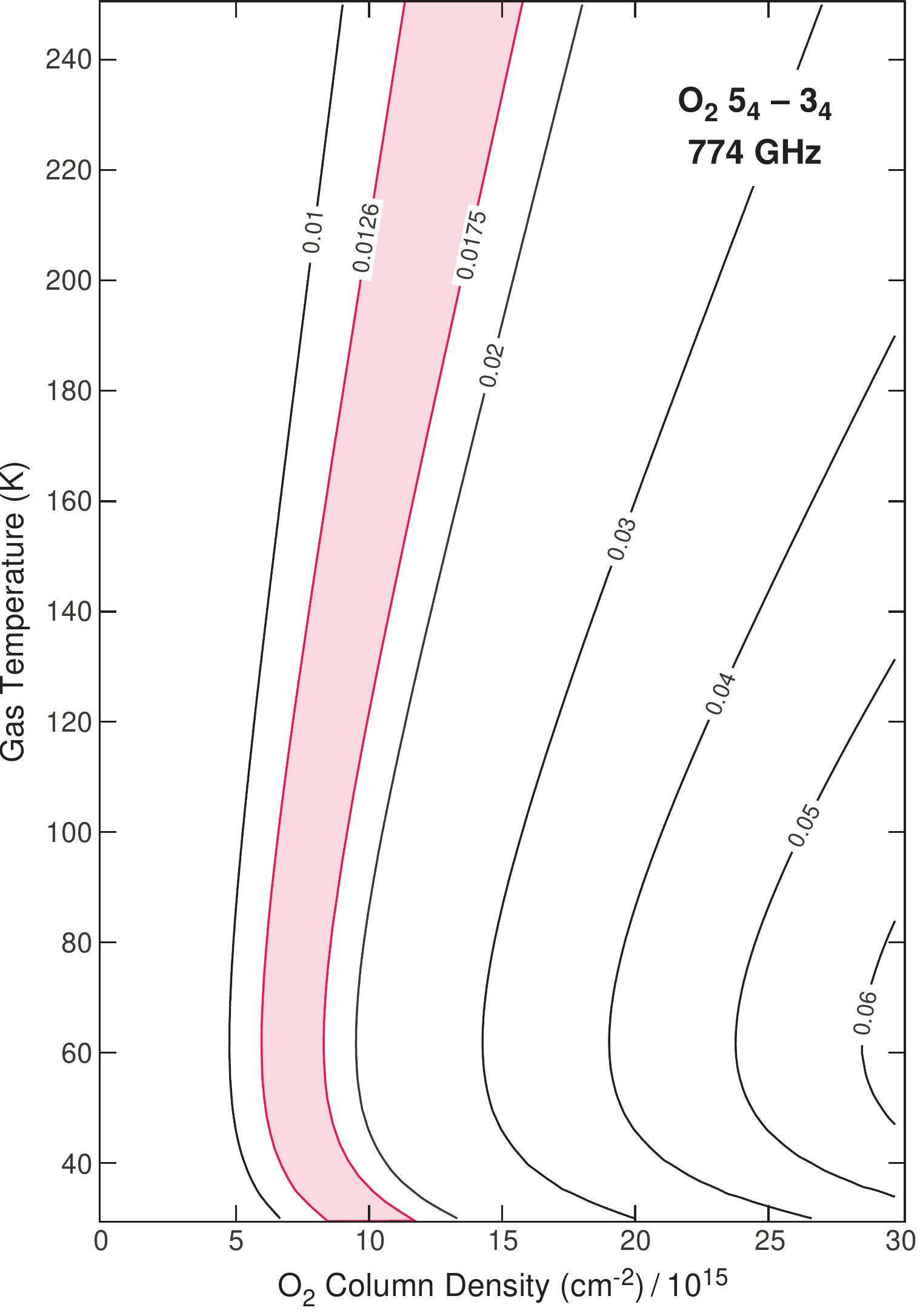}
\renewcommand{\baselinestretch}{0.99}
\vspace{3mm}
\caption{Contours of integrated antenna temperature, in K km s$^{-1}$, under a Gaussian line profile
versus line-of-sight \oxy\ column density and gas temperature assuming an aperture efficiency of 0.7. 
The shaded area bounds the range of upper limits to the \oxy\ 774 GHz integrated intensity assuming the
intrinsic \oxy\ line FWHM is 1.8\kms\ (0.0126 \kks) or 2.5\kms\ (0.0175 \kks).}
\renewcommand{\baselinestretch}{1.0}
\label{contours774}
\end{figure}

To determine the fractional population of the transition upper state, $f_u$, the excitation of
the lowest 36 levels of \oxy, corresponding to a maximum
upper-level temperature of 1141~K, was computed under the large 
velocity gradient (LVG) approximation.  The spontaneous decay rates are those of \cite{Drouin10}
and the collisional rate coefficients are those calculated by \cite{Lique10} for He\dash \oxy\ collisions,
multiplied by 1.37 to account for the different reduced mass when \mh\ is the collision partner. 
For molecular hydrogen densities $>\,$3$\,\times\,$10$^4$~\cmc, both the \tro\ and \trt\ transitions
are close to (or in) LTE and the values of $f_u$ depend essentially only on the temperature. 
Fig.~\ref{contours487} shows the resulting contours of integrated
antenna temperature for the \tro\ transition as functions of the total \oxy\ column density and 
gas temperature between 30 and 250~K.  Similarly, Fig.~\ref{contours774} shows the corresponding
results for the \trt\ transition.

Of the two \oxy\ lines searched for here, an examination of Figs.~\ref{contours487} and 
\ref{contours774} shows that our measured upper limits 
to the \trt\ 774 GHz integrated intensity 
place a more stringent limit on the maximum \oxy\ column density for \tgas$\:>\:$35$\:$K (and
comparable limits to that set by the 487~GHz line at \tgas$\:\sim\,$30$\:$K).  
Specifically, assuming the emission fills the beam, the total line-of-sight \oxy\ column density must be less than 
1.5$\,\times\,$10$^{16}$~\cms\ (3$\sigma$).   If the \oxy\ abundance peaks within the cooler
well-shielded gas, for which \tgas$\;\simlt\;$100$\:$K, 
the upper limit to the total \oxy\ column density is less than 1$\,\times\,$10$^{16}$~\cms\ (3$\sigma$).

\vspace{2mm}

\section{DISCUSSION}

\oxy\ is produced primarily through the gas-phase reaction O$\,+\,$OH$\,\rightarrow\,$\oxy$\,+\,$H
and is destroyed by photodissociation for the cloud depths of interest here.   Thus, the \oxy\ abundance
is expected to peak where the FUV field has been heavily attenuated and
where both the gas-phase O and OH abundances are high which,
in externally FUV-illuminated clouds, is predicted to
occur within a relatively narrow (i.e., a few \av\ deep) zone centered at an \av\ 
$\simlt\,$ 9 mag.~from the cloud surface
\citep[cf.][]{Hollenbach09}.  The proximity of  this zone to the surface and 
the range of depths over which the peak abundance occurs are governed by several 
important processes.  Near the cloud surface, where the FUV field is largely unattenuated,
the equilibrium \oxy\ abundance is low owing to a high photodissociation rate.
Beyond a few \av\ into the cloud, the FUV field is attenuated, the photodissociation rate reduced,
and a region of peak \oxy\ (and \water) abundance is attained.  

Within most clouds with \go$\,<\:$500, the path to \oxy\ formation is believed to start with the formation
of water ice, \waterice, on grains, which occurs when O atoms strike and stick to grains long enough to combine
with an accreted H atom
to form \ohice\ and then \waterice.   Within this region the FUV field remains strong enough to
photodesorb \water\ from the ice mantles and subsequently photodissociate these molecules, creating
sufficient gas-phase O and OH to produce \oxy\ by the gas-phase chemical reaction above.  
Deeper into the cloud (i.e., greater \av), the FUV
field is almost completely attenuated and the gas-phase OH and \water\ 
produced through the photodesorption and photodissociation of \waterice\ drops significantly;
most O atoms that then strike dust grains and form \waterice\ remain locked in ice as long as the
grain temperature is $\simlt\,$100$\:$K.  Over time ($\sim\,$10$^5$ years), 
this process greatly reduces the gas-phase atomic oxygen 
abundance and suppresses the formation and abundance of \oxy.  
%(The destruction of CO by cosmic-ray ionized He deep within clouds liberates some O atoms via the
%gas-phase reaction CO$\,+\,$He$^+ \!\rightarrow\,$C$^+\,+\,$O$\,+\,$He, but the effect on the
%total steady-state \oxy\ column density is negligible.  Likewise, the destruction of \oxy\ by He$^+$ is not important.)
Hence, in the model of \cite{Hollenbach09},
the steady-state abundance profile of \oxy\ (and \water) resembles an
elevated plateau that peaks at
an \av$\:\simlt\:$6 for gas with 
\nh$\:=\:$10$^4$\dash 10$^5$~\cmc\ and \go$\:\simlt\:$500.

\begin{figure}[t]
\centering
\vspace{-0.70in}
%$\!$\includegraphics[scale=0.605]{fig1.eps}
$\!$\includegraphics[scale=0.93]{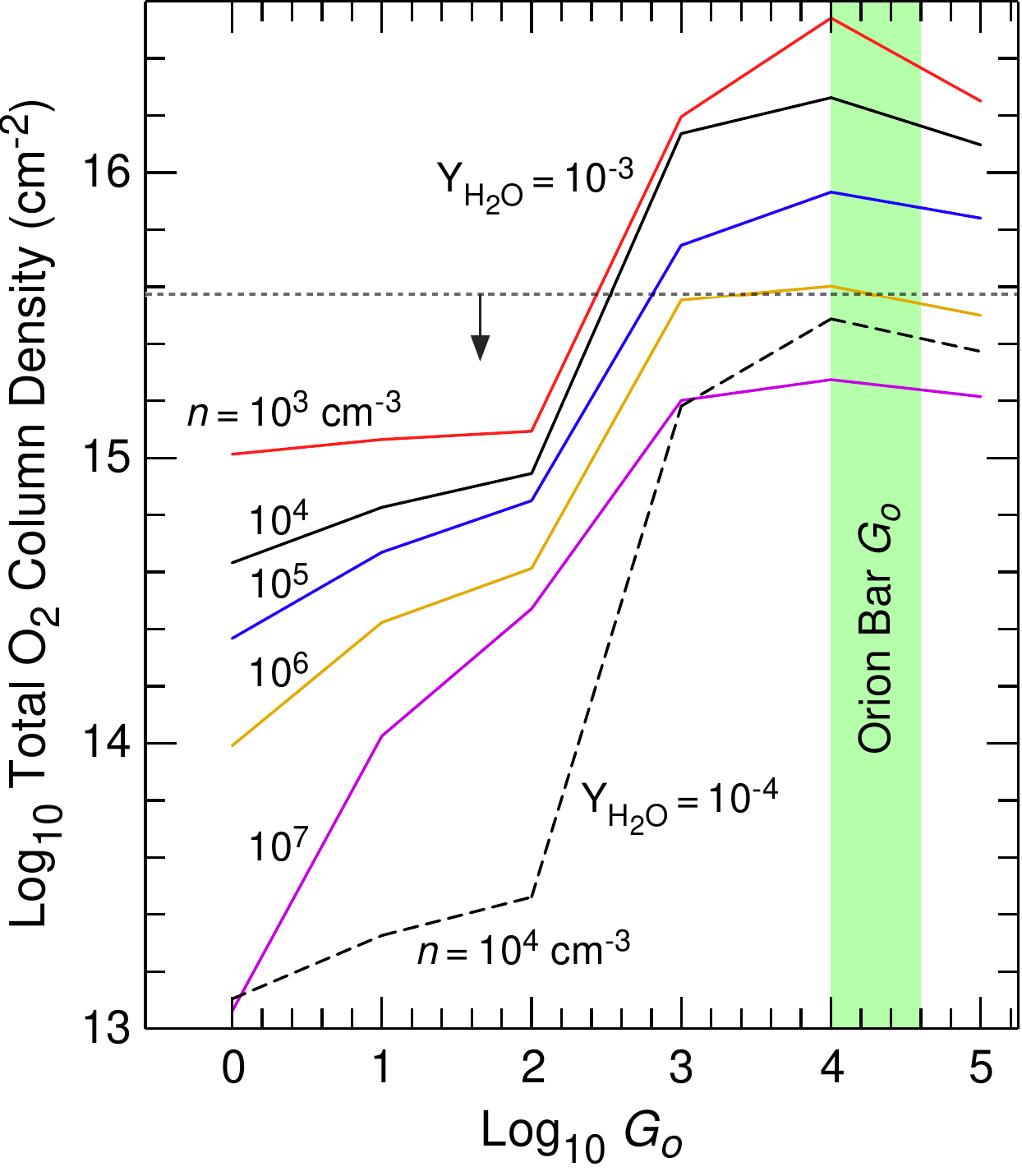}
\renewcommand{\baselinestretch}{0.99}
\vspace{3mm}
\caption{Predicted total \oxy\ column density perpendicular to the ionization front
as a function of \go\ and $n$(H$+$2\mh) for \water\ photodesorption
yields of 10$^{-3}$ (solid lines) and 10$^{-4}$ (dashed line).  The results shown assume a cloud thickness sufficient
to encompass the zone of peak abundance \citep[after][]{Hollenbach09}.  The range of \go\ that 
applies to the Orion Bar is shown in the shaded region.  The horizontal dotted line denotes
the upper limit to the \oxy\ column density established here, i.e., 1.5$\,\times\,$10$^{16}$~\cms, divided by 
a geometrical enhancement factor of 4.}
\renewcommand{\baselinestretch}{1.0}
\label{Go}
\end{figure}

For regions subject to a \go\ greater than $\sim\,$500, such as the Orion Bar, the scenario above is
altered and, for several reasons, the peak \oxy\ abundance is higher and occurs at a higher \av.  
First, the high FUV field absorbed at the cloud surface leads to a high infrared field that keeps
the grains warm, even deep within the cloud.  For \go$\:=\:$10$^4$, $T_{\rm gr}\:\approx\;$40$\:$K 
to \av$\,\simgt\,$8, resulting in a significant fraction of the O atoms being thermally desorbed 
from the grains before they can form \waterice\ and leading to an increase in O in the gas phase.
Second, the higher grain temperature also reduces the freezeout of such oxygen-bearing species
as OH and \oxy, further increasing the amount of elemental O in the gas phase.  Finally, the 
attenuated FUV flux at the higher  values of \av\ lowers the photodestruction rates, allowing \oxy\ to
survive to greater cloud depths.  The combined result of these effects is a peak \oxy\ abundance
about 3 times higher, and a total \oxy\ column density more than 10 times greater
than for comparably dense gas exposed to \go$\:\simlt\:$500.  This result is reflected in the
detailed calculations presented in \cite{Hollenbach09} and shown in Fig.~\ref{Go}, which is adapted
from their paper.  For this reason, the Orion Bar
was considered a promising source for our attempts to detect \oxy\ emission.

From Fig.~\ref{Go}, it would appear that the upper limits on the total \oxy\ column density established
here are not in  serious disagreement with the model predictions.  However, the results shown in Fig.~\ref{Go}
apply to a gas column perpendicular to the face of a planar cloud.  This is not the geometry of the Orion Bar,
which has often been described as an edge-on PDR, though its true structure has been the subject of 
some study and debate.  For example, based on millimeter and submillimeter line observations,
\cite{Hogerheijde95} and \cite{Jansen95} propose a model in which the Bar has a tilt angle, $\alpha$,
of $\sim\,$3\ddeg\ from
edge-on, resulting in an increase in the line-of-sight column density (beyond
what would be measured for a face-on geometry) by a factor 
(sin$\,\alpha$)$^{-1}$, or almost 20.  Alternately, \cite{Walmsley00} find
that a cylindrical model, in which the axis is in the plane of sky and the radius is 0.3$\:$pc, best reproduces
the observed spatial distribution of the fluorescent OI 1.317\um\ emission.  In this scenario,
the average geometrical enhancement of the line-of-sight depth into the Bar versus the face-on
depth is about 5.  Finally,
\cite{Neufeld06} find that a geometrical enhancement factor of
$\sim\,$4 is required to reconcile observed and predicted C$^+$ column densities. 

\begin{figure}[t]
\centering
\vspace{-0.20in}
%$\!$\includegraphics[scale=0.605]{fig1.eps}
$\!$\includegraphics[scale=0.79]{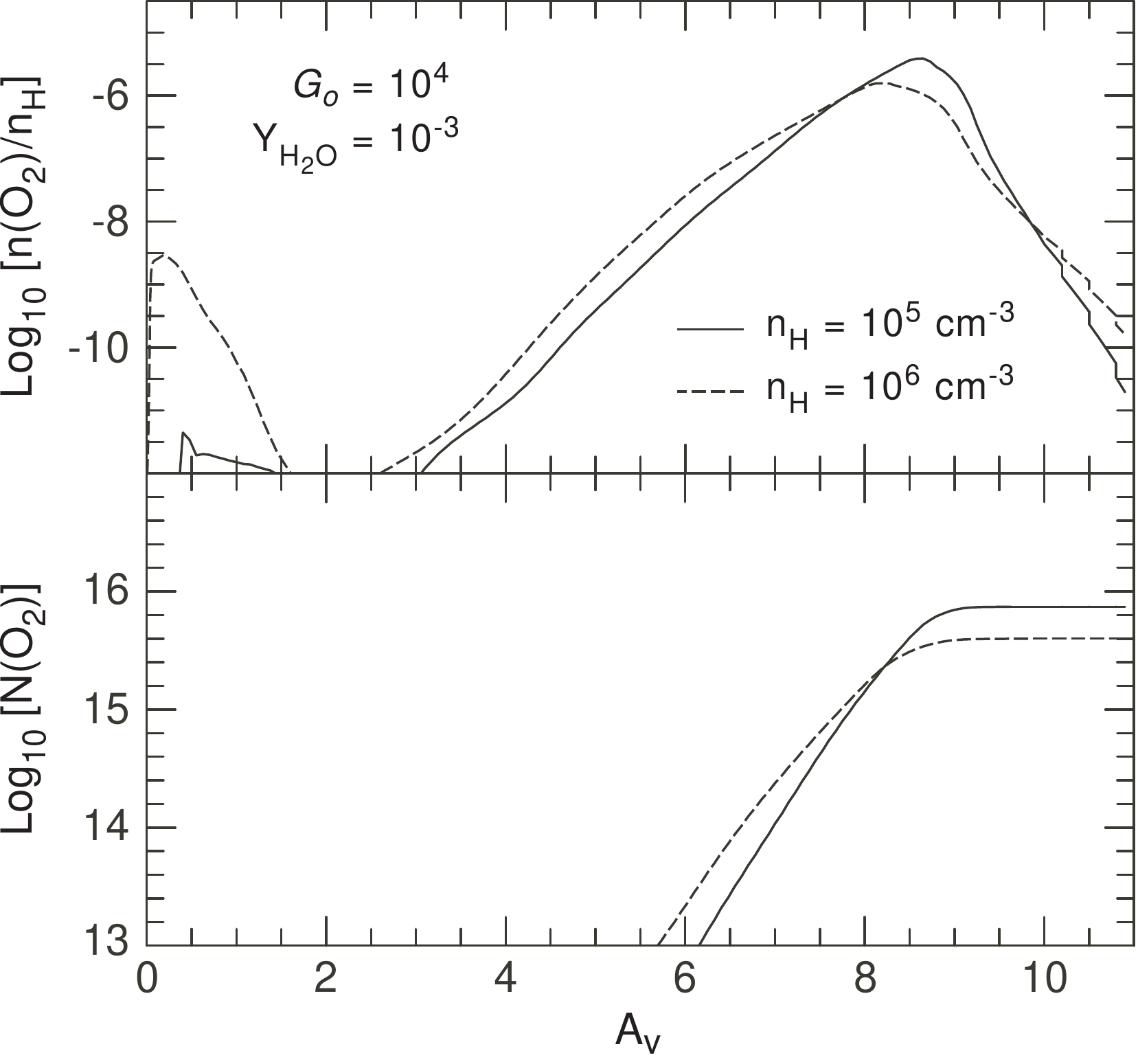}
\renewcommand{\baselinestretch}{0.99}
\vspace{5mm}
\caption{{\em Top panel:} Abundance of \oxy\ as a function of face-on depth into a cloud, measured
in \av,
for a cloud with $n_{\rm H} =\:n$(H$+$2H$_2$)$\:=\:$10$^5$~\cmc\ and 10$^6$~\cmc\ exposed
to a FUV field of \go$\,=\:$10$^4$.  This result was computed using the model described in 
\cite{Hollenbach09} assuming their ``standard" model parameters, except for those noted here.
An \water\ photodesorption yield of 10$^{-3}$ is assumed.
The gas and dust temperatures throughout the cloud are calculated self-consistently in the Hollenbach et al.~code,
which predicts a gas temperature of 33$\:$K, and a dust temperature of 42$\:$K, at the depth of the peak
\oxy\ abundance above.
{\em Bottom panel:} Cumulative face-on column density of \oxy\ integrated from the cloud surface 
to a given depth, in \av, for the abundance profile
shown in the top panel.}
\renewcommand{\baselinestretch}{1.0}
\label{Kaufman}
\end{figure}

The 3$\sigma$ upper limit to the {\em face-on} \oxy\ column density can thus be inferred from our line-of-sight
values to be 1.5\ti 10$^{16}\:$sin$\,\alpha$~\cms, or 1.0\ti 10$^{16}\:$sin$\,\alpha$~\cms\
for \tgas$\:\simlt\:$100$\:$K.  (We note that these upper limits are derived assuming the intrinsic
\oxy\ FWHM line width is 2.5~\kms; if the intrinsic width is closer to the lower end of the observed
range, i.e., 1.8$\:$\kms, the face-on \oxy\ column density upper limits are further reduced by a factor of 1.4.)
For gas densities $\simlt\,$10$^5$~\cmc, which applies to most of the gas in the Bar, this
is to be compared with a total predicted face-on \oxy\
column density of $\,\simgt\;$7$\,\times\,$10$^{15}$~\cms, as shown in 
Fig.~\ref{Go}, with most of this column occurring inside a layer of peak \oxy\ abundance with a width corresponding to approximately 2 magnitudes (see Fig.~\ref{Kaufman}), or a
linear size of $\sim\,$1.9\ti 10$^{16}$/$n_5$~cm, where $n_5\,=\;$\nh/[10$^5$~\cmc].
Viewed from a distance of 420~pc, this zone of peak \oxy\ emission would subtend 
3[(1/$n_5$~+~162.4$\,\ell\,$sin$\,\alpha$]\asec, where $\ell$ is the physical length of the Bar in parsecs.
For $\ell\,\simeq\;$0.6~pc \citep[cf.][]{Jansen95} and
$n_5 \simeq\:$1, $\alpha\,\simgt\:$6\ddeg\ would result in \oxy\ emission that fills the
{\em Herschel}/HIFI beam at 774~GHz, though a
minimum geometric enhancement factor of 4, derived from other observations,
suggests that $\alpha$ does not exceed 15\ddeg.  However, these tilt angles imply an upper limit to the 
face-on \oxy\ column density between 1.6\ti 10$^{15}$~\cms\ and 3.9\ti 10$^{15}$~\cms, which 
is below, and in some cases, significantly below that predicted by theory.

For $\ell\,\simeq\;$0.6~pc and $n_5 \simeq\:$1, but $\alpha <\;$6\ddeg,
the \oxy\ layer no longer fills the 774~GHz beam.
Although the peak \oxy\ column density within the beam will continue to increase
for angles less than 6\ddeg, the beam filling factor will decrease.  These two effects
offset exactly, and the beam-averaged \oxy\ column density will
remain the same for all tilt angles less than about 6\ddeg.  Since the \oxy\ emission is
optically thin, the line emission will likewise remain constant within the under-filled beam.
In this case, 
the geometrical enhancement factor would be $\sim\:$10, and the upper limit to the face-on \oxy\ 
column density remains below that predicted.
Therefore, we conclude that Bar geometry cannot account for the
discrepancy between theory and observations.

What, then, can account for the discrepancy?  The amount of \oxy\ produced in externally FUV-illuminated
dense gas depends on several factors, which we examine below:

\noindent {\em Thermal evaporation:}~As noted earlier, the dwell time of an O atom on a grain surface
can have a considerable effect on the \oxy\ abundance, particularly when this time becomes
less than the time to combine with an H atom on the surface.  The timescale for thermal
evaporation of an O atom is approximately 9$\,\times\,$10$^{-13}$ exp$\,$[800$\:$K / $T_{\rm gr}$] seconds,
where 800$\:$K is the adsorption energy of O to water-ice \citep{Hasegawa93} that applies to van der Waals binding
to a chemically saturated surface.   It is possible that the binding energy
is greater than 800$\:$K, which would increase the grain temperature, and thus the \go, required to
thermally desorb O atoms on short timescales and 
produce the jump in the total \oxy\ column density for 
\go$\,\simgt\:$500  seen in Fig.~\ref{Go}.  If, for example, the O adsorption energy was 1600$\:$K, grains
as warm as $\sim\:$42$\:$K -- the expected dust temperature at high \av\ in a 
\go$\,\simeq\,$10$^4$ field -- would, on average, retain their O atoms long enough to form \waterice,
thus delaying the \go$\:>\:$500 rise in \oxy\ column density seen in Fig.~\ref{Go} until \go$\:>\:$10$^4$.

\noindent {\em Photodesorption yield of H$_2$O from a grain surface, Y$_{H_2O}$:}~The abundance 
(and column density) of \oxy\
depends on the gas-phase abundance of O and OH, the latter being produced primarily through
the photodissociation of \water, much of which is either photodesorbed from grains or produced via
the dissociative recombination of gas-phase H$_3$O$^+$.  At high \go\ (and $T_{\rm gr} >\,$20$\:$K),
short O-atom dwell times on grains suppress the formation of \ohice\ and \waterice.
However, even though it is not formed on the grain surface in a high-\go\ environment, 
\water\ formed in the gas phase via H$_3$O$^+$ dissociative recombination will be depleted 
through freezeout onto grains and will remain locked in \waterice\ for as long as $T_{\rm gr} \simlt\,$100$\:$K.
Since the quantity of OH and \water\ returned to the gas phase as a consequence of \waterice\ photodesorption
scales with \ywat, the total \oxy\ column density likewise scales with \ywat,
as is seen in Fig.~\ref{Go}.  A value for \ywat\ less than 10$^{-3}$ would help to reconcile theory and
observation.  However, fits to the {\em SWAS} and {\em Odin} \water\ data \citep{Hollenbach09} as well as
theoretical simulations and laboratory measurements \citep{Andersson08, Arasa11, Westley95a, Westley95b, Oberg09} 
suggest, if anything, that the appropriate value of \ywat\ is greater than 10$^{-3}$.

\noindent {\em Grain cross-sectional area (per H):}~The
equilibrium \oxy\ abundance in the \av\ range of maximum \oxy\ abundance scales as (\ywat)$^{2\,}$\sigh,
where \sigh\ is the grain cross-sectional area per H nucleus.  
Therefore, lowering \sigh\ will decrease
the \oxy\ column density, bringing model and observation into closer agreement.
For an ``MRN"
\citep{Mathis77} grain size distribution $n_{\rm gr}(a)\propto a^{-3.5}$, where $a$ is the grain radius,
\sigh\ $\sim\:$2$\,\times\,$10$^{-21}$~cm$^2$ for an assumed gas-to-dust mass ratio of 100 with 
grains ranging in radii 
between a minimum, \rmin, of 20$\:$\AA\ and a maximum, \rmax, of
2500$\:$\AA\ (the standard value in \citealt{Hollenbach09}).
Grains with \rmin$\,<\;$20$\,$\AA\ will be
cleared of ice mantles by single photon heating or cosmic rays and, thus, are not significant ice reservoirs.
Because \sigh$\:\propto\:$($a_{\rm min} \cdot a_{\rm max})^{\;-0.5}$, 
in order to lower the value of \sigh\ while preserving the total mass in grains, either or both \rmin\ and \rmax\
must {\em increase}, such as
through coagulation.   For example, a reduction in \sigh, and thus the face-on \oxy\ column density,
by at least a factor of 2 could be achieved if the minimum grain radius were to increase to $\simgt\,$80$\:$\AA.

Alternately, the buildup of an ice mantle, which can increase the radius of grains by as much as 
$\sim\,$50$\:$\AA, will increase the value of \sigh.  For values of \go\ of $\sim\,$10$^4$ applicable
to the Orion Bar, grain temperatures are expected to be $\sim\,$40$\:$K, which is 
high enough to inhibit ice formation via surface reactions
(absent a higher O adsorption energy); however, water formed in the gas phase via the reaction 
H$_3$O$^+ +\,e^-\!\rightarrow\;$\water$\,+\,$H can still freeze out and form an ice mantle.
Toward Orion, there is evidence for a departure from the assumed gas-to-dust mass ratio of 100,
which is consistent with the buildup of ice mantles \citep[see, for example,][]{Goldsmith97}.
In addition, there is evidence for  a deficiency in small grains and for grain growth, 
possibly due to radiation pressure, the preferential evaporation of small grains, and coagulation 
\citep[e.g.,][]{Cesarsky00, Pellegrini09, Shaw09}.
The net effect of lowering \sigh\ through these processes, and
increasing \sigh\ through the accumulation of an ice mantle, is unclear in a high-\go\ environment
like the Orion Bar.

\noindent {\em Beam position:}~For an interclump \mh\ density between 5$\,\times\,$10$^4$~\cmc\ and 
5$\,\times\,$10$^5$~\cmc\ and \go$\:=\:$10$^4$, the peak \oxy\ abundance is predicted to occur 
at a face-on depth into the cloud corresponding to an \av$\;\sim\;$8 (see Fig.~\ref{Kaufman}).  
Thus, the linear distance from the \av$\:=\:$0 surface,
which we assume is the prominent ionization front, to the depth of peak \oxy\ abundance is 
$\sim\:$7.6$\,\times\,$10$^{21}$/\nh~cm.  For an assumed distance of 420~pc, the angular
separation between the ionization front and the position of peak \oxy\
abundance (and column density) is then $\simeq\:$1.5$\,$\av/[\nh/10$^5$] 
arcseconds, where \av\ is the face-on depth of the \oxy\ peak abundance in magnitudes.  
Thus, an interclump \mh\ density of 10$^5$~\cmc\ should produce \oxy\
emission that peaks $\sim\:$12\asec\ from the ionization front and close to the center 
of the observed sky positions (see Fig.~\ref{finderchart}).  
However, if the interclump density is more than
a factor of 2 different from 10$^5$~\cmc\ -- values that remain within the range of density estimates 
for the interclump medium -- then the peak \oxy\ abundance is predicted to fall to either side
of the observed beam center position.

Finally, we note that the inferred {\em peak} line-of-sight \mh\ column density, $N$(\mh), 
applicable to the interclump medium toward the Orion Bar
is estimated to be 6.5$\,\times\,$10$^{22}$~\cms\ \citep{Hogerheijde95}.  If the geometrical
enhancement factor is $\simgt\,$10, as would be the case for a tilt angle $\simlt\:$5.5\ddeg,
this would imply a face-on \mh\ column density of $\simlt\;$6.5$\times\,$10$^{21}$~\cms,
corresponding to a total \av\ through the Bar of about 7.  
If the face-on extinction through the Orion Bar is indeed this low, then the
attenuation of the \go$\,\sim\,$10$^4$ field is not sufficient to allow \oxy\ to reach its peak abundance 
and the total \oxy\ column density will be less than predicted by \cite{Hollenbach09}, whose 
total column densities are based upon cloud depths corresponding to \av$\,\geq\,$10.  This is
illustrated in Fig.~\ref{Kaufman}, which shows both the profile of \oxy\ abundance versus \av\
and the cumulative \oxy\ column density to a given \av, computed using the model described 
in \cite{Hollenbach09} for the conditions appropriate to the Bar
interclump medium.  At a depth corresponding to an \av\ of 7, the predicted 
face-on \oxy\ column density remains $<\:$3$\,\times\,$10$^{14}$~\cms, well below the limits set here.

The clumps known to exist
within the Bar do possess higher \mh\ densities (i.e., 10$^6$\dash 10$^7$~\cmc) and
column densities \citep[i.e., $>\,$10$^{23}$~\cms;][]{Lis03} and would provide the necessary FUV shielding
to allow \oxy\ to reach its full predicted abundance.  Such conditions help to reconcile observation
and theory in two ways.  First, as shown in Fig.~\ref{Go}, the predicted total \oxy\ column
densities {\em decrease} with higher \mh\ densities.   Thus, the total \oxy\ column 
density is predicted to be lower if the \oxy\ emission arises primarily from within the dense clumps rather
than the surrounding lower density interclump medium.
Second, interferometric observations indicate that the dense clumps within the Bar typically subtend angles of between
4\asec\ and 8\asec\ \citep[see, for example,][]{Lis03}, 
and thus provide a natural explanation for why the beam filling factor of \oxy\ emission could be
less than unity.  However, whether the correct explanation for what we observe is that \oxy\ emission
originates preferentially within the dense clumps, and is suppressed within the \av$\,\simlt\,$7 interclump
medium, and with both gas components governed by the processes described in
\cite{Hollenbach09}, will
depend on how well this model reproduces the wealth of new
lines being detected toward the Orion Bar by {\em Herschel}.

\vspace{2mm}

\section{SUMMARY}

1.$\,$~We have conducted a search for \oxy\ toward the Orion Bar, carrying out deep integrations around
the frequencies of the $\rm N_J =\;$\tro\ and \trt\ transitions at 487~GHz and 774~GHz, respectively.  Neither
line was detected, but sufficiently sensitive limits on their integrated intensities were obtained to test current 
models of molecular gas exposed to high fluxes of FUV radiation -- i.e., \go$\,\sim\,$10$^4$.
In particular, we infer a total face-on \oxy\ column density of $\simlt\:$4$\,\times\,$10$^{15}$~\cms,
assuming a Bar geometry in which the line-of-sight depth is more than 4 times greater than its
face-on dimension.  This column density is at least 2 times less than that predicted by the model
of \cite{Hollenbach09} for
the densities, temperatures, and \go\ appropriate to the Orion Bar.

2.$\,$~The discrepancy between the model predictions and our observations would be reduced, if not eliminated,
if the adsorption energy of atomic oxygen to wate-ice were greater than 800$\,$K, and possibly as high as 1600$\:$K.
A lower value for the
photodesorption yield for \water\ would help, but is not supported by fits to other astronomical data or
recent theoretical calculations and laboratory measurements.
A lower grain cross-sectional area per H, such as might occur through grain coagulation, radiation
pressure, or the preferential destruction of small grains, 
would lower the \oxy\ column density, but it is unclear whether these grain properties apply within
the Orion Bar.

3.$\,$~If the total face-on depth of the interclump medium within the Orion Bar corresponds to 
an \av$\,\simlt\,$7, then photodissociation will reduce the \oxy\ column density to values below our
detection limit.  Clumps embedded within the Bar would offer 
sufficient shielding to enable the buildup of higher \oxy\ abundances and column densities in accord
with model predictions,
while the small filling factor of these clumps
would reduce the \oxy\ line flux to levels consistent with our upper limits.  

4.$\,$~If the total face-on depth of the interclump medium within the Orion Bar corresponds to 
an \av$\,>\,$8, it remains possible that most of the \oxy\ emission may have been missed.
In particular, since the gas density affects
the angular separation between the ionization front and the face-on depth into the 
Bar at which the \oxy\ abundance is predicted to peak, 
interclump \mh\ densities much different than the assumed value of 10$^5$~\cmc\ could 
result in the position of peak \oxy\ abundance and column density occurring to either the northwest
or southeast of the position we selected.

Only further modeling,
including predictions for other species, can establish which, if any, of the above possibilities is most likely
to resolve the present puzzle.

\vspace{0.3in}

HIFI has been designed and built by a consortium of institutes and university
departments from across Europe, Canada and the United States under the leadership of
SRON Netherlands Institute for Space Research, Groningen, The Netherlands, and with
major contributions from Germany, France and the US. Consortium members are: Canada:
CSA, U. Waterloo; France: CESR, LAB, LERMA, IRAM; Germany: KOSMA, MPIfR,
MPS; Ireland, NUI Maynooth; Italy: ASI, IFSI-INAF, Osservatorio Astrofisico di Arcetri-INAF;
Netherlands: SRON, TUD; Poland: CAMK, CBK; Spain: Observatorio Astron\'{o}mico
Nacional (IGN), Centro de Astrobiolog\'{a} (CSIC-INTA). Sweden: Chalmers University of
Technology - MC2, RSS \& GARD; Onsala Space Observatory; Swedish National Space
Board, Stockholm University - Stockholm Observatory; Switzerland: ETH Zurich, FHNW;
USA: Caltech, JPL, NHSC.  We also acknowledge the effort that went
into making critical spectroscopic data available through the
Jet Propulsion Laboratory Molecular Spectroscopy Data Base
(http://spec.jpl.nasa.gov/), the Cologne Database for Molecular
Spectroscopy (http://www.astro.uni-koeln.de/cdms/ and \citealt{Muller05})
and the Leiden Atomic and Molecular Database
(http://www.strw.leidenuniv.nl/$\sim$moldata/ and \citealt{Schoier05}).  Finally, it is a pleasure
to acknowledge useful discussions with Dr.~Edwin Bergin.

Support for this work was provided by NASA through an award issued by JPL/Caltech.

\clearpage

\bibliographystyle{plainnat}

%\clearpage

\begin{sidewaystable}[htdp]
%\caption{default}
\begin{center}
TABLE~1.~Summary of Observations \\*[1.5mm]
\begin{tabular}{lcccccccc} \hline
\rule{0mm}{6mm} & & & &  & \multicolumn{4}{c}{Gaussian Fit Parameters} \\*[0.3mm] \cline{6-9}
\rule{0mm}{6mm} & & Rest & ~Observing~ & Integration & T$_{\rm A}^{\;*}$ & LSR Line & & Integrated \\*[-0.6mm]
Species & Transition & Frequency$^{1}$ & Mode$^{\, 2}$ & ~Time~ & Amplitude & Center & FWHM & Intensity \\*[0.3mm]
 & & (GHz) & & (hrs) & (K) & (km s$^{-1}$) & ~(km s$^{-1}$)~ & ~(K-km s$^{-1}$)~ \\*[1.3mm] \hline
\rule{0mm}{6mm} H$_2$Cl$^+$ & $J =\,$1$_{11}$\dash 0$_{00}$ & & & & & & & \\
 & F$ =\,$3/2\dash 3/2 & 485.413 & sc & 1.16 & ~~0.055 & 10.56 & 2.47 & 0.15 \\
 & F$ =\,$5/2\dash 3/2 & 485.418 & sc & 1.16 & ~~0.076 & 10.56 & 2.47 & 0.20 \\
 & F$ =\,$1/2\dash 3/2 & 485.421 & sc & 1.16 & ~~0.030 & 10.57 & 2.47 & 0.08 \\*[3.2mm]
~SO$^+$ & $J =\,$21/2\dash 19/2  & ~~~486.837~~~ & sc & 1.85 & ~~0.029 & 10.77 & 2.28 & 0.07 \\
 & $\Omega =\,$1/2, $\ell =\,$e  & & & & & & & \\*[3.2mm]
~SO$^+$ & $J =\,$21/2\dash 19/2  & 487.212 & sc & 1.85 & ~~0.027 & 10.99 & 1.86 & 0.05 \\
 & $\Omega =\,$1/2, $\ell =\,$f  & & & & & & & \\*[3.2mm]
~O$_2$ & 3$_3$\dash 1$_2$ & 487.249 & sc & 1.85 & $\leq\,$0.008$^{\, 3}$ & \dash & \dash & \dash \\*[3.2mm]
~CS & $J =\,$10\dash 9 & 489.751 & sc & 0.46 & 0.46 & 10.58 & 1.78 &  0.87 \\*[3.2mm]
~\ico & $J =\,$7\dash 6 & 771.184 & sp & 1.15 & 27.04~~ & 10.67 & 2.24 & 64.48~~~\\*[3.2mm]
~O$_2$ & 5$_4$\dash 3$_4$ & 773.840 & sc, sp & 10.91~~ & $\leq\,$0.007$^{\, 3}$ & \dash & \dash & \dash \\*[3.2mm]
~C$_2$H & N$\,=\,$9\dash 8 & 785.802 & sc, sp & 10.91~~ & 0.34 & 10.76 & 2.35 & 0.84 \\
 & $J =\,$19/2\dash 17/2  & & & & & & & \\
 & F$ =\,$9\dash 8  & & & & & & & \\*[3.2mm]
~C$_2$H & N$\,=\,$9\dash 8 & 785.865 & sc, sp & 10.91~~ & 0.30 & 10.77 & 2.35 & 0.75 \\
 & $J =\,$17/2\dash 15/2  & & & & & & & \\
 & F$ =\,$9\dash 8  & & & & & & & \\*[3.2mm]
~C$^{17}$O & $J =\,$7\dash 6 & 786.281 & sc, sp & 10.91~~ & 1.19 & 10.62 & 1.76 & 2.23 \\*[1.9mm] \hline
\end{tabular}
\end{center}
\vspace{-6.7mm}
\phantom{0}\hspace{9mm}$^{1}$~NRAO-recommended rest frequency.~~$^{2}$~sc:~spectral scan observation;~~sp:~single point observation. \\
\phantom{0}\hspace{7mm}$^{3}$~3$\sigma$ upper limit.
\label{tobs}
\end{sidewaystable}%

\clearpage

\clearpage

\clearpage

\clearpage

\clearpage

\clearpage

\clearpage

\end{document}